# A LEO-Based Solar-Shade System to Mitigate Global Warming

Rahul Suresh
Bangalore, India, rahulsuresh1987@gmail.com

Andrew Meulenberg
NAv6 Center of Excellence, Universiti Sains Malaysia, Penang
781-577-1594, mules333@gmail.com

The development of a Low-Earth-Orbit (LEO) based solar-shade system, as part of a technically- and financially-viable multipurpose system to provide long-term solutions to global warming and the energy crisis is discussed. The proposed solar-shade/power system would be enabled by the development of a previously-proposed less-expensive, environment-friendly, space-elevator system to lift mass into space. The solar shades, even during their early deployment and growth in LEO would provide benefits such as reduction of space-debris and depletion of the Van-Allen radiation belts.

The Terrestrial temperature profile has been approximated for each latitudinal zone with a one-dimensional model. A shade ring at an altitude of 2000 - 4000 kms, consisting of thin-film mega panels totaling up to 4% of the earth's surface area (to block ~1% of insolation), is proposed. The effects of such near-polar rings on the global temperature pattern has been examined using the simple model. Specific emphasis has been laid on this effect in the Polar Regions. One such proposed ring could reduce the peak summer temperature of the Polar Regions (80-90° latitude) by almost 3°C. The "tilting" of the solar-shades, to reduce its cooling effect at the poles and to increase it in the near-polar regions is recommended.

## INTRODUCTION

The development of a Low-Earth-Orbit (LEO) based solar-shade system, as a precursor to a global solar-power system, and its effects on global temperature patterns to mitigate global warming is discussed in this paper.

Space-based Geo-engineering systems - Overview:

Several Geoengineering projects,[1] like Solar-Power Beaming from GEO,[2,3] sequestering of $CO_2$,[4] sulfur sun-shades,[5] Fresnel lenses at L1,[6] etc., have been proposed for many years. Most of these projects, such as solar-shades and solar-power satellite systems, when implemented as stand-alone systems, are not (and may never become) technically or economically feasible. Nevertheless, we have recently shown that the shade/solar-power system makes economic sense when developed as a part of a multi-purpose system such as the "LEO ARCHIPELAGO™"- a multipurpose Low-Earth-Orbit–based communications/mass-lifter/solar-shade/space-power system.[7,8,9] The financial return from the communications system would not only attract initial investment, but also would sustain the growth of subsequent stages of the system.

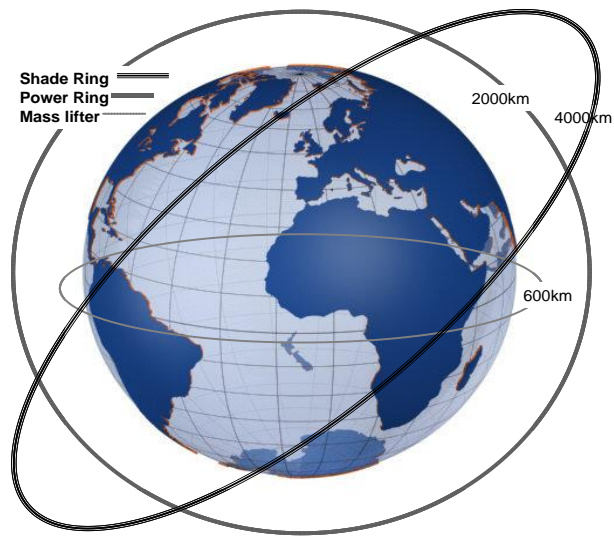

Fig. 1: Aspects of the ring system (see text) proposed to solve a number of man's problems and to become the basis for "stepping stones" into space.

Figure 1 indicates the principal elements of the proposed LEO ARCHIPELAGO™ ring system. It consists of: an equatorial ring at ~600km for the mass lifting and communications (space and terrestrial) portion of the system; a sun synchronous "power ring" (at ~2000 km) for collection, conversion, storage, and distribution of the solar energy available 24 hours a day in this orbit; and a shade ring (at >3000 km) with orbital inclination and orientation determined by cooling and lighting requirements on the earth.

Basics of the proposed shade system:

Earth receives approximately 176,000TW of power from the sun. Even if a small fraction of this power (say 1%) is blocked, it would play a significant role in countering the global temperature rise. This would involve deployment of many million square kilometers of thin-film reflectors into space. Undertaking a project of this magnitude as a stand-alone system, using existing launching technologies such as rockets would be an economic and environmental disaster.

The system's success will depend, at least in part, on the timely development of carbon-nanotube (CNT)[10] or colossal carbon tube (CCT)[11] tethers as a means of enabling an efficient and cost-effective mass-lifter and circum-Terra ring system. International cooperation would be required to organize and fund such a major project that would affect so many countries and special interests. Naturally, one would also expect many technical, environmental impact, and legal hurdles. Generating a positive public opinion about such undertakings is crucial to the creation of a strong political will. A recent survey[12] in India has shown that 90% of respondents think of space-based systems to solve global warming and energy crisis issues as a major potential benefit to be derived from the global space program.

The proposed mass-lifter system, "Sling-on-a-Ring™,"[8] (capable of transporting payloads into space in an environment-friendly and economical manner), would have many competing demands on it once in place. One straight-forward application would be the launch and deployment of large-area, low-mass, thin sheets of opaque or reflective material (such as aluminized Kapton[13] into rings at high-LEO. This shade-ring structure would block or reflect away a fraction of the total sunlight reaching the Earth, hence causing a temperature drop on the surface of the Earth.

The large-area, thin films capable of shadowing around 1% of the earth's surface, if installed using rocket launch, would constitute a massive burden on the environment and world finances (in terms of rocket effluent and costs). If, on the other hand, this heavy system can be more efficiently launched into space, by use of the mass-lifter, the proposal is much more attractive. On reaching the intended construction site in equatorial orbit (greater than 1500 km above the surface of the earth) these rolls of micro-sheeting (2-6 micrometers thick, or thicker - but lighter, if the low-density CCT materials can be space qualified) can be automatically deployed along and across the circum-Terra rings envisioned. This process is much less costly and technically less demanding than the deployment of solar collectors and could still bring immediate financial return in the form of equivalent "carbon credits."[14] These solar shades, initially in equatorial construction "rings," would be collectively raised into higher orbits and tilted into sun-synchronous, near-polar, slant or tilt orbits. During the initial construction phase, the large area films deployed in lower orbits will steadily deplete the space-debris and trapped radiation in those regions. As construction proceeds and the rings are expanded out to higher orbits, these regions will become cleared also. Thus, more and more near-earth space will be made habitable and useful for man's growth into space.

Cooling the Polar Regions is more important than cooling the equator (owing to the enormity caused by the melting of polar ice).[15] Therefore, to benefit from even the early stages of construction, later shade rings could be constructed directly in the higher-altitude and higher-slant orbits. This later development would be dependent on the ability to lift (or inject) mass directly into these orbits and the ability of the earlier shades to reduce space-debris and radiation levels in the Van Allen belts sufficiently to allow human workers and robots to operate in these "new" regions. This early expansion into near-polar orbits could play a major role in preventing the melting of polar sea ice and of glaciers. The melting of sea ice does not raise sea level; but, it does result in higher solar absorptance in those regions. The melting of glaciers does raise sea level. In this manner, the shades will be "buying time" while a more massive, technologically challenging, and expensive, solar-power system, SPS, is being deployed to replace the fossil fuels causing the crisis.

This paper primarily deals with the study of effects on global temperature patterns of a shade-ring in polar/near-polar orbit at an altitude of 1500 to 4000 kms, blocking ~1% of the insolation. Other aspects of the LEO-ARCHIPELAGO, such as: impact of new materials on the space-elevator development timeline, laser-power beaming, etc., are discussed in another session.[16]

## SYSTEM MODELS

In this section, a brief description of the zero- and one-dimensional energy balance models[17,18] to study the temperature profiles of different latitudinal zones is provided. These models, modified to account for solar shades, are used to identify the scale of the proposed undertaking. For the sake of simplicity (and a 1-D model), the shade has been assumed to be a ring in near-polar orbit with a surface area perhaps exceeding 4% of the earth's surface and at an altitude of 2-4000 kms. It blocks approximately 1% of the incoming solar radiation from hitting the earth. The energy-balance models, with deployment of the shade system, assume the reflection of sunlight by a solar-shade is properly accounted for by an appropriate reduction in global and local insolation values,[19] respectively.

### Zero dimensional model

The simplest method of considering the climate system of the Earth, and indeed of any planet, is in terms of its global energy balance. Viewing the Earth from outside, one observes an amount of radiation input which is balanced (in the long term) by an amount of radiation output.

The details of a 0-D model are discussed in Appendix A. If our shade blocks about 1% of the energy coming to the Earth, the global average temperature would be reduced by ~ $1.1^{\circ}C$.

### One dimensional model

Though favorable results are shown by the 0-D model, it does not provide us all the information required to study the system's effect on global warming mitigation. For example, it does not indicate the effect of shadowing on each latitudinal zone. It is particularly important for us to study the seasonal effect of shade in the near-polar regions because cooling of these regions in summer mitigates a major consequence of global warming. Hence, the next logical step puts spatial resolution into our model. We do this by breaking each hemisphere into 9 latitude regions of $10^{\circ}$ each.

Details of a 1-D model are discussed in Appendix B. The 1-D model takes into account solar flux and albedo variations for different latitudinal zones to attain thermal equilibrium. Nevertheless, the effects of ocean currents, latent heat in the rain cycles, and latent heat in ocean and land ice in equalizing regional temperatures are not always considered. The actual data in Fig. 2 indicates that the northern hemisphere shows greater temperature extremes because a larger portion of its total area is covered by land and ice, hence it experiences a lesser moderating effect of oceans and their currents.

Figure 2 indicates the actual temperature profile of the earth, from published data.[20] Note the differences between, and variations in, land and ocean temperatures. This becomes important in the modeling of both the regional temperatures and the effects of shades on them. In particular, the ocean temperatures are "pinned" near zero Celsius by the latent heat of sea ice.[21] In looking at the effects of solar shades in these critical regions, this latent-heat effect must be eliminated to get a true picture of the benefits.

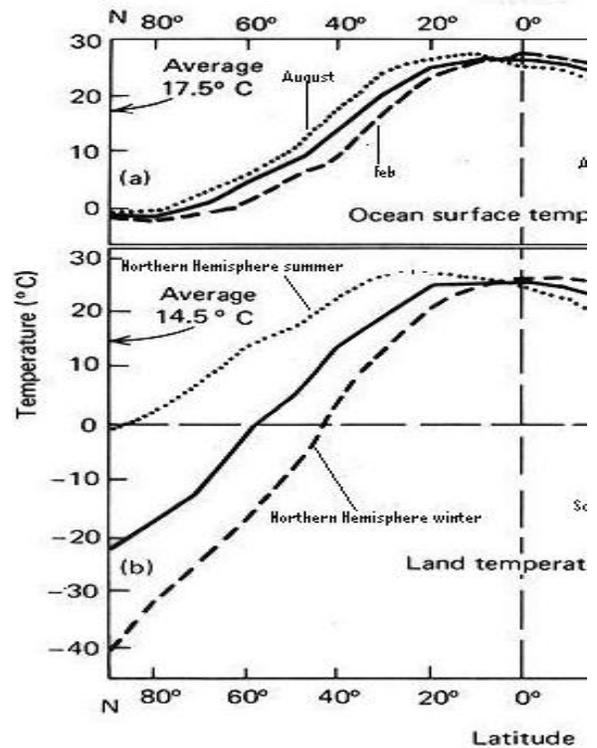

Fig 2: Global Temperature profiles based on data.[20]

Of major concern, but for different reasons, is the melting of ice from the landmasses and the ocean surfaces. The ice melting on Greenland and Antarctica both can contribute to a rise in sea level. But, melting of the sea ice, while not contributing to that danger, is a major contributor to global warming by simply decreasing the albedo over large expanses of ocean. Thus, we must look at the temperatures near freezing on both sea and land masses. Figures 2 indicate the nature of the problem even today. An increase in temperature of only a few degrees in the polar regions will greatly accelerate the melting of the Greenland ice mass and the polar sea ice. The massive sea-ice shelves about Antarctica are already breaking up. This will reduce the protection from the warming ocean that they presently afford the glaciers. Again, the interrelated effects of global warming and ocean rise proceed hand-in-hand.

To match the modeled results with existing data and desired changes, we consider a model where the daily-average solar flux and albedo[22] both have latitude dependence ($S_i$ and $\alpha_i$, where i runs from 1 to 9 in each hemisphere indicating the latitude band). Also, if a latitudinal band is colder or warmer than the average regional temperatures, heat will flow into/out-of the band. This time and spatial averaging addresses a portion of the thermal inertia observed from heat/cold stored in the earth, sea, and ice. To simplify the model,[23] we assume that the dominant portion of this heat flow depends linearly on the temperature difference between each region and its average temperature $T_{Avg}(i)$, in other words it is: $F\times(T_i-T_{Avg})$, where F is the *Heat Transport Coefficient (F=3.80 W m$^{-2}$ $^o$C$^{-1}$)*

Since the primary interest of our study is to observe the effects of a shade system (i.e. the drop in global and regional temperatures due to the presence of shades), and the equalizing effect of ocean currents would be similar with/without the shades, we have limited our system analysis to a 1-D model. However, higher-order models would certainly provide additional details that may be vital in fully understanding the system performance and hence should be an area of focus in the future.

We have modified the existing 1-D models to specifically calculate global temperature profiles during solstice because the effect of our system on the Polar Regions, during the summer, is crucial to the efficacy of this approach. The algorithm used for 1-D energy-balance models, both with and without the presence of the proposed shade system(s) has been discussed in Appendix C. A table containing various calculated data values has also been provided.

Effects of Shades

The orientation of the shade ring(s) would be sun-aligned so that it would cast a shadow 24 hours a day. Beyond that condition, there are a number of variables that can be used to optimize for the various goals. Fig. 3 (with the reader in the sun position) indicates 3 options that fulfill different goals: polar orbit - for maximal global cooling; slant orbit - for maximal summer-time cooling; and tilt orbit - for maximal cooling of a particular latitude range.

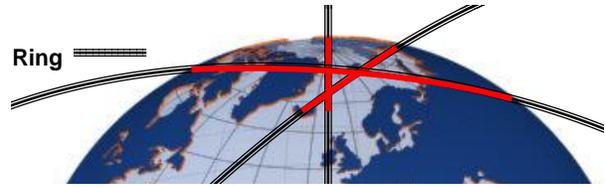

Fig. 3: Three options (of many) for solar-shade rings: polar orbit (vertical); slant orbit; and tilt orbit (horizontal). From the sun position, the length of a tilt-orbit shadow cast (in red) over the critical polar regions (Latitude > 65$^o$) is increased > 3x, relative to the polar-orbit shadow

Polar-orbit shading is the simplest for which to calculate the effect. Using the summertime insolation values ($s_i$) in Table B (Appendix B), one is able to determine the projection area of the shadow cast by the ring over the arctic region. Subsequent division of the shadow area by the latitudinal area gives the relative reduction of insolation for that region. Since the shadow extends over both sides of the northern-most latitudinal zones during this period (when the two rings are fully illuminated), the insolation values are high (despite the still-oblique angle from the sun). Furthermore, since the area of these extreme rings is small relative to that of the equatorial latitudinal rings, the relative reduction of insolation from the ring shadow is high. Therefore, the effect on temperature from the shadow of a polar-orbiting shadow ring is greatest for the polar region.

Fig. 4 provides details of this shadowing effect on the average daily temperature for the northern latitudes during summer solstice. The uppermost curve (solid with no markers) is the baseline solar input and radiation output temperature-balance curve. The dotted curve shows the polar-shade impact as greatest in the 80 - 90$^o$ N region. The lower solid curve includes a correction for thermal-inertia and latent-heat effects. The dashed curves are for the two solid curves with the tilt-orbit shade effect included to show optimization of the shade for cooling in the 75$^o$ band.. Note that, even without latent-heat effects, the temperature curve flattens in the 75$^o$ and 85$^o$ bands as a result of their full illumination during this period (and thus their increased insolation values, $s_i$ in Table B).

Shading from a ring in a tilt orbit (e.g., the horizontal ring in Fig. 3) mainly limits the effect to only a couple of the latitudinal regions (75$^o$ and 65$^o$ bands). Since the polar-most region (85$^o$) has more thermal "cold" stored and is further from any warming ocean currents, its ice can perhaps survive without being shaded, if the adjacent regions can be more highly shaded as a consequence.

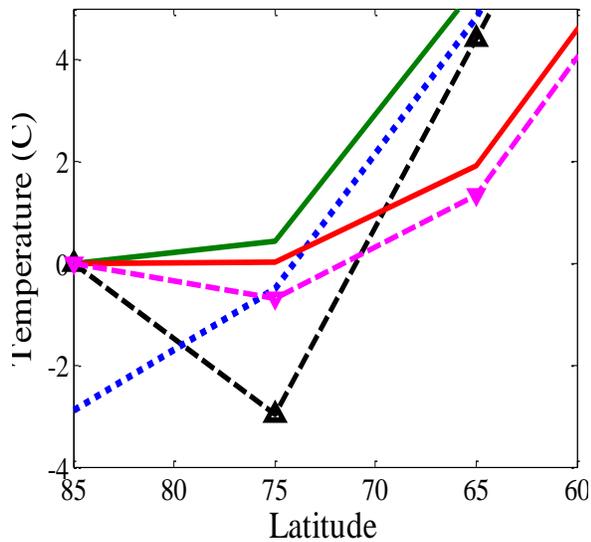

Fig. 4 Effects of solar shades on terrestrial temperatures for two different models (with and without thermal inertia, see text) and on two different shade configurations (polar and tilt - dotted and dashed respectively).

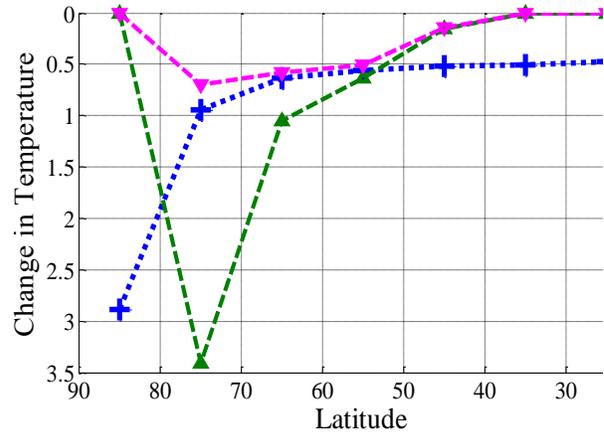

Fig. 5. Shade-ring effects (degrees C cooling as a function of Latitude) for different shade ring orientation (triangle-up for tilt ring, "plus" markers for polar ring) and for a different thermal mass and latent-heat model (triangle-down).

The upper dashed curve, with the triangle-up markers, in Fig. 4 indicates the effects of the tilt-orbit shade in Fig. 3. Note that there is no impact on the polar-most region (nor on the regions south of ~55°N, not shown). However, there is a large reduction in temperature (>3°), over the larger area in the regions needed to preserve Greenland's glaciers and the Arctic's summer-ice fields.

The actual change in temperature resulting from the solar shades will not be as great as indicated above. The thermal mass and the latent heat of the ice fields will "buffer" the effect. This buffering is seen in comparing the upper and lower sets of curves in Fig.4. The lower set is the same as the unshaded and the tilt-orbit shaded results of the upper set except that a "time-lag" between increased solar input and the actual temperature increase (resulting from the winter cold and the latent heat of the melting ice) is approximated. Details of the effect of this buffering in the ice fields is represented in Fig. 5.

The triangle-up marked curve in Fig. 5 directly compares the cooling effect (ΔT) of shading for the a tilt-orbit (triangle up) with that of a polar-shade ring ("plus" markers). The greater lowering of temperature over a larger area region for the tilt orbit is an indication of the versatility of the ring-shade system. Note that only the polar-shade ring is still cooling at the southern latitudes.

The triangle-down marked curve in Fig. 5 is also for the tilt-orbit, but with some thermal buffering effects included. The buffering (primarily from the sea ice) keeps the surface temperatures pinned near zero Celsius (as seen in Fig. 4) and prevents loss of ice rather than lowering the temperature in response to the shading. This does not mean that the shade ring is not being effective. It means that the actual change in temperature is not the critical point. Prevention of ice loss by shadowing is the primary goal, because that keeps the local ocean from being a good solar absorber.

The polar-orbit shade ring (plus-marked curve) is less affected by such buffering because its greatest effect is in colder regions that (as yet) are not quite in the ice-water-transition region. (The thickness of the sea ice in this extreme polar region is sufficient to act as a thermal barrier to prevent much additional ice growth or to allow the temperature of the ocean to influence the surface temperatures.) The surface area of the 75° band is 3x that of the 85° band. So, even though global cooling from the tilt-orbit shade ring may not be as great as that from a polar-orbit shade ring, the change in temperature distribution is different; cooling in the critical region is greater and more efficient by being directed.

Further refinements of shade orientation

The orientation of the shade rings can be further refined (e.g., seasonally adjusted) depending on the shade size (e.g., during its construction) and on the terrestrial requirements. As an example mentioned above, sea ice has a major impact on increasing the albedo relative to water. Therefore, it is desirable to use the shade rings to shadow the ocean near the ice sheets, in terms of maximizing the direct cooling from the shade ring. (Since the water has low albedo, the effect of reducing insolation is greater there than over the ice sheet itself.) However, the greatest effect comes if the shade ring can prevent the melting of the sea ice over large areas. If this can be accomplished, it becomes a major "multiplier" of the shade-ring effect. Thus, it would be more efficient, in the long run, to shade the ice than it would be to reduce heating of the sea water. This also has global implications. If a 1% shade can prevent sea ice from melting, then the resulting increased albedo can reduce warming in that region as effectively as would a 20% shade. Globally, when shading the ice sheets near both poles, this could greatly increase the efficiency of the shade ring.

If a full-sized tilt-orbit shade ring produces a 10% reduction of insolation in a latitudinal band, a much smaller one could have an equivalent effect if it simply prevented the ice from melting over an equivalent area. Therefore, if started soon enough (i.e., in years, not in the season, before the polar regions get warmed much beyond their present levels), then a small tilt-orbit shade ring oriented to $65^o$ rather than to $75^o$ could be as effective as a much larger one constructed later and could be sufficient to prevent many of the major consequences of global warming. Considering the size and cost of the shade rings, an early investment (that might forestall the need for a larger system) could be very wise.

Similar considerations apply to Antarctica. Stabilizing the coastal ice sheets becomes as important as directly reducing any melting of the glaciers. The interaction of the different systems and the potential for delaying the melting of ice in the coastal regions means that the ability to seasonally "adjust" shadowing is a control mode that makes the shade rings such an important asset. Furthermore, other "multiplier effects" have been identified[24] whereby a small change in local insolation can have a disproportionate effect on the regional temperatures. An obvious one is the shading of regions with the highest solar absorption during the equinox period, when the ice sheets are not threatened. This would include the large ocean regions and, in particular, the cloudless ocean regions. Not so obvious (but if the models are correct) is that a positive feedback mechanism exists that would multiply this shading by an order of magnitude. Thus, even during the growth stage of the shade ring, it could have a impact on the global temperature that would far exceed the actual percent reduction in the earth's insolation.

ADDITIONAL APPLICATIONS OF THE SHADE RING(S)

Solar Energy Conversion and Storage

The application of the shade rings as radiation and space-debris absorbers, in addition to their main purpose, has already been mentioned. However, there are other applications that ultimately could become as important as the primary goal, once the global warming crisis is past. The ring itself is the basis for a new civilization (remember, if it is 4% of the earth's surface, it is larger than any country). Its mechanical inertia is a storage medium of immense capacity and high efficiency. The panels, if oriented, can be inefficient, but very powerful, drivers for converting solar pressure into mechanical (orbital) energy. This mechanical energy can then be efficiently converted to electrical energy via electrodynamic tethers. Furthermore, the large temperature gradients available between the sunlit and shaded regions of the ring are a further source of "free" energy that will always be available locally on the ring. With ready access to raw materials of the moon and asteroid belts, it meets all of the needs for an advanced civilization.

Twilight Ring

The Shade Rings have been primarily considered as "passive" elements in man's reshaping the earth and moving into space. Nevertheless, the use as solar "mills" is definitely an active application with immense potential payoff beyond the initial goals. The shade ring also becomes a potential aesthetic addition to man's environment. If the earthside of the rings are made to reflect light only in the visible spectrum, then we have visible rings as lovely as Saturn's, visibly much brighter than the moon, but introducing relatively little heat. In fact, the solar reflection from the earthside of a tilt-orbit shade-ring could illuminate the winter sky and eliminate the need for early evening and dawn lighting for cities and highways of the world (mainly in the northern hemisphere). These "twilight" rings would introduce only a small amount of the total solar energy incident on them; but, they would reduce the need for terrestrial power (with its attendant $CO_2$ or radioactive waste) that is presently being used for such lighting. Furthermore, the heat that **is** introduced by this application is only in the winter hemisphere, where it would cause no harm relative to the alternative (terrestrial electricity for lighting).

Costs:

There is a "rule of thumb" for leisure items. If you have to ask how much it costs, you probably can't afford it. If the shade rings were purely for comfort, as in air-conditioning buildings, the cost would be too high - even if it were the only way of doing it. On the other hand, cooling the earth (at least partially) is a necessity from an economic and, perhaps, even from a survival-of-nations viewpoint. Therefore, the ability to afford a project is not an option. The choices must be made and expenses incurred. It is a matter of what you get (or hope to get) for your money.

As a stand-alone system, based on rocket-launched materials, the shade-ring concept is a non-starter. As a portion of a larger system, and based on non-rocket mass-lifting capabilities, it is a solution and an investment at a fraction of the cost expected for $CO_2$ sequestration. It does not directly solve all the $CO_2$ problems created by fossil fuels (such as acid rain and ocean acidification, etc.), however, it leads to, and supports, technologies and systems that can solve these problems. It is not just burying the problem; it is solving the problem by helping to create a system with the potential for immense profits and ultimate survival of the human race.

Is the technology up to the task? From the time that the LEO-Ring System was first proposed last year, technology has already surpassed the expected 10-year milestones. There are no foreseen "show stoppers" (other than not starting). Politically, some people prefer warfare as a means of making profits and controlling population. $CO_2$ sequestration is a similar approach - spend trillions of dollars that are not voluntarily contributed by taxpayers for a product that has no intrinsic value and no hope for any return except for maintaining the status quo, stimulation of a portion of the economy, and the profits that are derived within specific sectors.

Are there alternatives to the shade-ring system with the same starting assumptions? If there are, then they need to be explored. With the assumption of a functional and inexpensive mass-lifter, the option of a giant sun shade at L1 becomes a possibility. The total mass involved in an equivalent (controllable) system is reduced by perhaps a factor of 4. There are other problems in such a system. However, they may no longer be show-stoppers, if the mass-lifter technology is assumed. The technological and economic challenges and benefits of both systems need to be compared. There may not be a single answer, and it is undesirable to depend on a single solution. The point is, there are now "drivers" for systems that could not be seriously explored in the past. There are new technologies that did not exist when such options were explored earlier. We cannot afford not to explore them again in the light of new developments.

## SUMMARY

A global cooling system has been proposed and explored in the context of countering global warming and of developing a Low-Earth-Orbit civilization. Different aspects of the warming problem have been explored. The main thrust here has been to determine what kind of temperature effects can be expected from a given solar shade size and with some of the options available. In particular, the possibility of "multipliers" to enhance the effects of local shading have been mentioned. This impacts the timeliness of intervention via shade-rings. With that basis, a rough comparison can now be made with alternatives to achieve similar goals. As a starting point, this paper has addressed most of the reservations that the recent Royal Society Report[25] has expressed about the near-term feasibility of space-based techniques for solar control.

While the proposed system seems to be science fiction, or at least a far-future event, the required technology and materials are real (well beyond the "proof of concept" stage). System development should begin today. As far as its application to global warming, the sooner begun, the lower the total system requirements and costs.

## ACKNOWLEDGEMENT

This work is supported in part by HiPi Consulting, New Market, MD, USA, by the Science for Humanity Trust, Bangalore, 560094, India, and by the Science for Humanity Trust, Inc, Tucker, GA, USA.

## Appendix A: Zero-dimensional analysis of Earth's energy balance

In an energy balance model, the main goal is to account for all heat flows in ($P_{Gain}$) and out ($P_{Loss}$) of the system.[26] If these are balanced ($P_{Gain}=P_{Loss}$), the system will be in a steady state and the system will be at a constant temperature. If the heat flows are not balanced, the temperature of the system will change. The *Solar Constant* (S) = energy arriving (during a 1 second period on a 1 square-meter area oriented perpendicular to the sun's rays) from the sun at the upper atmosphere. (*The annual average value of the solar constant is taken as S=1370 W/m²*.)

A fraction of the sun's radiation is immediately reflected or reradiated back into space, either from the atmosphere, clouds, or the earth's surface. The net amount of solar radiation arriving on a 1 m² area (perpendicular to sun) of the earth's surface is $S(1-\alpha)$, *where α is the albedo of the earth.*

From the point of view of the sun, the earth appears to be a disk with a radius R, so the total amount of power absorbed by the whole earth is the product of the arriving solar radiation times the area of a disk the size of the earth minus the power reflected:

$$P_{Gain} = \pi R^2 S(1-\alpha).$$

*By Stephan's law,* $P_{LOSS} = 4 \pi R^2 \sigma T^4$, *where σ is Stefan's constant and T is the average global temperature in Kelvin.* Assuming $P_{Gain} = P_{LOSS}$,

$$\pi R^2 S(1-\alpha) = 4 \pi R^2 \sigma T^4 \Rightarrow T = [S(1-\alpha)/4\sigma]^{1/4}.$$

Considering parameters to compute average global temperature, *S = 1370 W/m², Average albedo = 0.32, σ = 5.67E-8 W/m²K⁴*. Substituting the above values in the equation, we estimate the average global temperature to be around **253K**, which is much lower than the actual temperature (about **285K**). The major reason for this difference is the *Greenhouse Effect* in the atmosphere, which prevents much of the reradiated energy from escaping to space.

Since all of our temperatures are within about +/-30°C, it is possible to re-write the Stefan-Boltzmann equation using the binomial expansion: $(1+x)^n \sim (1+nx)$ if x is much less than 1. Therefore, $T_K^4 = (T_0 + T)^4 = T_0^4(1 + T/T_0)^4 \sim T_0^4(1 + 4T/T_0)$. This allows us to re-write the Stefan-Boltzmann equation as $P_{Loss} = (4\pi R^2)(A + B*T)$. The greenhouse effect can be included by modifying the values used for A and B. Therefore,

$$\pi R^2 S(1-\alpha) = 4 \pi R^2 (A + B*T) \text{ and}$$

$$T = [S(1-\alpha)/4 - A]/B.$$

## Appendix B: One-dimensional analysis of Earth's energy balance

The next logical step is to allow spatial resolution into our model. We do this by breaking each hemisphere into 9 latitude regions of 10° each. The general geometry is shown in Fig. B-1.

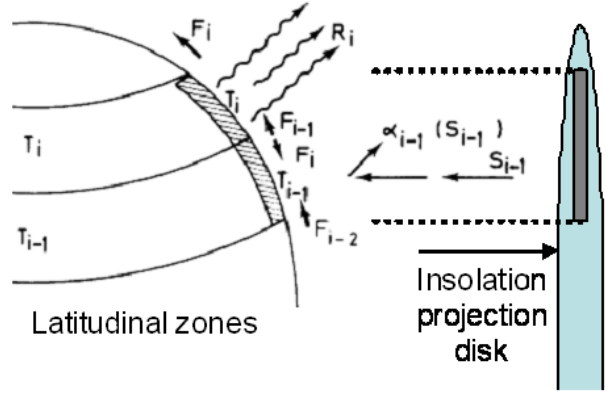

Fig. B-1. Terms for analysis. Based on *A Climate Modeling Primer*, A. Henderson-Sellers and K. McGuffie, Wiley, pg. 58(1987)]

In the model used, the solar flux and albedo both have a latitude dependence ($S_i$ and albedo, where i runs from 1 to 9 indicating the latitude band). Also, if a latitudinal band is colder or warmer than that of the adjacent (spatially or temporally) regions, heat will flow into/out-of the regions. We are looking for differences with respect to shadowing and season, not as a function of time. Therefore, to 1st approximation, we assume that this heat flow depends linearly on the temperature difference between the region and its annual average, in other words:

$$F \times (T_i - T_{Avg}).$$

$$C_E \, dT/dt = P_{GAIN} - P_{LOSS}$$

$$P_{GAIN} = S_i (1 - \alpha_i)/4$$

$$P_{LOSS} = A + B*T_i + F*(T_i - T_{avg})$$

$$\Rightarrow T_i = [S_i(1-\alpha_i) + F*T_{Avg} - A]/(B+F)$$

- $S_i$: Solar Flux in Latitude Band " i " (W m⁻²) The product of S/4 (the average global solar constant) times the insolation $s_i$.

- $s_i$: Solar Insolation : The **fraction** of incident flux on each latitudinal band (seasonally dependant).

- $\alpha_i$ : Albedo in Latitude Band i (seasonally dependant, but to a lesser degree). The albedo of ice/snow is much higher than that of land/water.

- F: Heat Transport Coefficient (F=3.80 W m⁻² °C⁻¹)

- $T_{Avg}(i)$ : Annual Average Temperature for bands

| Latitude band | Albedo ($\alpha_i$) (archival data) | | s(i) insolation | | | Area(i) / earth Area | Zonal Temp (Annual average) | Modeled Temperatures | |
|---|---|---|---|---|---|---|---|---|---|
| | Summer (N) | Winter (N) | Summer (N) | equinox | Winter (N) | | | Summer (N) | Winter (N) |
| 80-90° N | 0.44 | 0.75 | 1.46 | 0.421 | 0.00 | 0.0076 | -20 | -1.6 | -35 |
| 70-80 | 0.49 | 0.83 | 1.46 | 0.562 | 0.00 | 0.023 | -18 | -1.2 | -33 |
| 60-70 | 0.39 | 0.78 | 1.39 | 0.696 | 0.00 | 0.037 | -14 | 0.0 | -30 |
| 50-60 | 0.37 | 0.56 | 1.31 | 0.820 | 0.07 | 0.050 | -3 | 4.1 | -20 |
| 40-50 | 0.32 | 0.46 | 1.37 | 0.929 | 0.29 | 0.062 | 3 | 11 | -11 |
| 30-40 | 0.26 | 0.37 | 1.31 | 1.021 | 0.47 | 0.071 | 12 | 19 | -1.2 |
| 20-30 | 0.25 | 0.3 | 1.31 | 1.093 | 0.66 | 0.079 | 16 | 22 | 2.0 |
| 10-20 N | 0.2 | 0.27 | 1.23 | 1.141 | 0.88 | 0.084 | 19 | 22 | 10 |
| 0-10° N | 0.24 | 0.26 | 1.17 | 1.166 | 1.02 | 0.087 | 21 | 22 | 16 |
| 0-10° S | 0.25 | 0.21 | 1.02 | 1.166 | 1.17 | 0.087 | 21 | 16 | 23 |
| 10-20 S | 0.24 | 0.21 | 0.88 | 1.141 | 1.23 | 0.084 | 19 | 11 | 24 |
| 20-30 | 0.25 | 0.23 | 0.66 | 1.093 | 1.31 | 0.079 | 15 | 2.5 | 23 |
| 30-40 | 0.3 | 0.27 | 0.47 | 1.021 | 1.31 | 0.071 | 12 | -0.6 | 18 |
| 40-50 | 0.39 | 0.33 | 0.29 | 0.929 | 1.37 | 0.062 | 7 | -7 | 13 |
| 50-60 | 0.47 | 0.41 | 0.07 | 0.820 | 1.31 | 0.050 | 1 | -17 | 3.8 |
| 60-70 | 0.77 | 0.46 | 0.00 | 0.696 | 1.39 | 0.037 | -5 | -23 | 0.6 |
| 70-80 | 0.88 | 0.61 | 0.00 | 0.562 | 1.46 | 0.023 | -19 | -34 | -9 |
| 80-90° S | 0.8 | 0.58 | 0.00 | 0.421 | 1.46 | 0.0076 | -44 | -54 | -25 |

Table B. Input parameters pertaining to our 1-D analysis and modeled results.

A & B: Coefficients expressing Infrared Radiation Loss (A=204 W m$^{-2}$ and B=2.17 W m$^{-2}$ °C$^{-1}$)

$C_E$: Heat Capacity ($C_E = 2.08 \times 10^8$ J/m$^2$ °C)[26]

The following algorithm, and Table B, was used to consider the effect of shades.

1) The earth is divided into 18 different lat. zones

2) The albedo a(i) and solar influx value s(i)*S/4 for each of the 18 latitudinal zones are obtained from online archives (values can differ significantly)[26].

3) The surface area of each zone is calculated,
   Area(i) = Area$_{i, i+1}$ = 2 π R$^2$ (sin φ$_i$ - sin φ$_{i+1}$ )

4) Using the formula:
   $T_i = [S_i(1- \alpha_i) + F \cdot T_{avg}(i) - A]/(B+F)$

   The seasonal temperature of each latitudinal zone is calculated. From this, the global verage is calculated using the weighted mean,

   $T_{avesum} = \Sigma$ Area (i)* $T_i$ /Area of Earth.

5) Now a shade in polar orbit is considered such that it shadows 1 % of Earth's area. Except at equinox, the decrease in solar flux in each latitudinal zone would **not** be, $\Delta S_i = 0.01 * S_i$ (which implies a uniform change in the incident flux). The shadow band is a ratio of the effective areas of the ring shade to that of the corresponding latitudinal zones. See Appendix C for details.

6) With shades, the insolation in each zone may change: s$_i$_new(i) = s$_i$ - shadow$_i$. However, since S is multiplied by s$_i$ (to get S$_i$, in item 4) the shadow effect might be best defined in terms of S. Therefore, the usage could be S$_i$_new = S$_i$ - {equivalent shadow on insolation disk = (disk-shadow area/disk area) * S.

Appendix C: One-dimensional analysis of a Shade Ring on earth's energy balance

The insolation s$_i$ term is the means of keeping track of the differences between portions of one surface area of a disk (πr$^2$) and corresponding portions of the whole surface area of a sphere (4πr$^2$). The whole-area difference is a factor of ¼ x. The difference between the center of the disk and its projection onto the sphere is 1x. The difference between the central band of the disk and its projection onto the sphere (the equator) is 1/π. The difference between the top edge of the disk and its projection onto the sphere is 1/π$^2$ (~1/10) x. Therefore, ratio of sunlight falling on a very thin equatorial band to the average solar-power density is s$_i$' = (S/π) / (S/4) = 4/π = 1.27. For a broader band (corresponding to a latitudinal zone), the ratio is given by the equatorial value (in Table B) s$_9$ = 1.167. Similarly, the ratio for the polar zone $s_1$ equals 0.421 instead of $4/\pi^2 = 0.406$. The solar-energy removal, by shadows from a shade ring, must be considered in this context.

Determination of the temperature effects of different shade rings can be simplified in several cases. In the equinox period, the shadow cast by a polar-orbit shade ring is a vertical line between the north and south poles. The area on the insolation disk that will cast a 100km wide, 1108 km long, shadow across the 85° band of the earth's surface is a 100 x 97 km rectangle (Table C). The shadow cast in this zone by a 1% shade ring (~$1.1*10^5$ km$^2$) will cover >5% of the illuminated area (~$2*10^6$ km$^2$) of this band. Since insolation is based on the full-band area, the reduction of insolation will be ~2.5%. (The insolation reduction in the much-larger-area equatorial bands will be less that 1, so that the total (global) insolation reduction averages to 1%.

During the summer solstice, the North Pole is shifted toward the sun (by ~ 23.5°). Using the equinox and summertime insolation values in Table B, one is able to determine the projection area onto the insolation disk of the shadow cast by a polar ring during this period. Insolation of the 85° band, $s_1$, during this period is about 3.5x that of the equinox values, since it is fully, and more directly, exposed to sunlight during summer solstice. However, since the polar-orbit shade gives a vertical shadow and the shade is perpetually perpendicular to the solar flux (reducing the curvature factor of $\pi$ - somewhat less because of the large shadow size relative to the 85° band), the seasonal insolation increase of 3.5x for the band translates into an increase for a vertical shadow of nearly $\pi$ times 3.5x = 11x (see ~9x in Table C and below). Furthermore, the albedo values of the polar regions are lower during this period (Table B), because the sunlight is incident at a much-less oblique angle.

Calculation: The polar-orbit shadow projection onto the insolation "disk" is lengthened by the cosine of the reduced angles (Table C column 3). (An increase in area of the insolation disk projection of a latitude ring means that more solar power is incident on that ring.)

Since the 85° and 75° rings are now fully exposed to sunlight, the shadow projection is lengthened further and both sides of these rings must be considered. One side of the 85° ring is centered at ~62°; the other side at ~72°. The total increase in projection area of the 85° ring shadow, from this seasonal shift, is ~9x (calculated). The shadow will cover about 5% of the illuminated surface area of the band. Since the illuminated area of this band is also doubled during solstice, the total impact of a polar-orbit shade ring on the band is about 2.5%, if other effects are not considered (see Fig. 4).

One side of the 75° ring is centered at 52°; the other at ~82°. The calculated increase in total projection onto the insolation disk area from this seasonal shift is >3x. Again, the factor of 2, because of full illumination would seem appropriate, However, the curvature of the earth is now more important and (as seen in Fig. 4) a value of ~1% (rather than 1.5%) is measured. The other rings in the upper latitudes are simply lengthened (and those in the lower latitudes correspondingly shortened). The measured values are
less than the calculated values because the earth's image used for the measurements is from near-earth space rather than the sun's viewpoint and use of the center angle of a band is a less-accurate approximation for the near-polar zones. Earth's curvature, image parallax, and the viewer's position, relative to the equatorial plane, alters the measured values of the projection. Nevertheless, the 20% accuracy of this measurement is adequate to demonstrate the shade-ring effect.

Tilt-orbit ring: The shadow of the tilt-orbit ring is near-horizontal rather than vertical (as for the polar-orbit ring). Thus, the vertical projection of the shade-ring onto the insolation disk does not change with the season (this is now determined by the shade ring, not the latitudinal band). Since we have chosen to tilt the ring to maximize the horizontal shadow-length across a particular latitude band at solstice, that is the determinant of the increase in solar flux onto that band. The projection from the 85° band onto the insolation disk at equinox is nearly square (~97x100km), so we have a means of measuring the shade-ring width. We assume that the vertical size does not change with the season and that the area of the projection can therefore be determined solely from the horizontal length of the shadow times the known shade-ring width.

| Zone (or band) angle(s) | Projection of shadow onto insol. "disk" at **equinox** (km) | Solstice zone angle(s) | Projection of shadow onto insol. "disk" at **solstice** (km) | Ratio Solstice/ Equinox (calc.) | Ratio Solstice/ Equinox (meas.) | Length of in-band tilt-shadow / band height |
|---|---|---|---|---|---|---|
| 85 | 97 | 61.5 + 71.5 | 884 | 9.1 | 8.8 | 0.0 |
| 75 | 288 | 51.5 + 81.5 | 857 | 3.0 | 2.6 | 10.5 |
| 65 | 470 | 41.5(+88/3) | 833 (+13) | 1.8 | 1.6 | 3 |
| 55 | 638 | 31.5 | 948 | 1.5 | 1.2 | 1.7 |
| 45 | 786 | 21.5 | 1035 | 1.3 | 1.0 | 0.4 |

Table C. Data used in Figs. 4 and 5 from an orbit tilted further than that in Fig. 3

Measurement: Since the projection for a 1% area polar shade at equinox is $0.01*s_i$, the length of a polar-orbit-shade projection **at equinox** can be used as a calibration point for each band. Thus, the measured lengths, at each band of the polar-orbit shadow at solstice (e.g., from Fig. 3), is just a multiple of the equinox values for those bands. The multiple can be measured directly (within the 20% range mentioned for the present mapping of the globe) or calculated from the inverse cosine of the mid-band angle at solstice.